\documentclass[a4, 10pt]{article}
\pdfoutput=1

\usepackage{amsmath, amssymb, bm}
\usepackage[top=1in, bottom=1.2in, left=0.5in, right=0.5in]{geometry}
\usepackage[]{graphicx}
\usepackage[font=scriptsize,labelfont=bf]{caption}
\usepackage[caption=false]{subfig}
\usepackage{textcomp}
\usepackage{authblk}
\usepackage{hyperref}

\newcommand{\bs}{\boldsymbol}

\begin{document}

\title{\Large{\bf{Non linear quantum electrodynamics, longitudinal photon state and cosmological distances}}}
\author{Damian Ejlli}

\affil{\emph{\normalsize{School of Physics and Astronomy, Cardiff University, The Parade, Cardiff CF24 3AA, United Kingdom}}}

\date{}

\maketitle

\begin{abstract}

In this work, I study the non-linear QED effects to the Maxwell equations due to the interaction of the electromagnetic waves with an external magnetic field with arbitrary direction with respect to the direction of propagation of the electromagnetic waves. I show that in the case when the external magnetic field has a longitudinal component with respect to the electromagnetic wave direction of propagation, a longitudinal photon state is generated in a magnetized vacuum. The mixing of transverse and longitudinal photon states causes a decrease in the group velocity with respect to the velocity of electromagnetic waves in a vacuum. In a cosmological context, as far as a cosmic magnetic field is concerned, the decrease of the group velocity of any form of electromagnetic radiation would cause changes in several distance measures in cosmology. The most affected of these distance measures turn out to be the comoving distance and the luminosity distance. The  absolute change in distance measures are usually large but the relative changes are very small.
\end{abstract}

\section{Introduction}

Quantum electrodynamics (QED) is among the most established theories in physics that have been tested multiple times and that agrees with experiments to high accuracy. The most important predictions of QED are non-linear corrections to Maxwell equations in flat spacetime. Indeed, QED predicts that self-interaction of the electromagnetic field or its interaction with external prescribed electromagnetic fields, would lead to non-linear corrections to the Maxwell equations. These non-linear corrections make it possible the appearance of birefringence and dichroism effects in the vacuum that in principle can be tested in a laboratory. These new effects were initially predicted and calculated by Euler and Heisenberg\cite{Heisenberg:1935qt} by using the Dirac theory of the positron which in the Language of Feynman diagrams correspond to the one-loop correction to the Maxwell equations for a spinor field. The predictions of the Euler-Heisenberg theory were found in the case of low energy interacting electromagnetic fields and the extension to arbitrary energies was done by Karplus and Neumann\cite{Karplus:1950zz} where standard QED calculations were employed. 

The consequences of the Euler-Heisenberg theory have been studied by several authors and especially the non-linear corrections to the Maxwell equations in the case of interaction of electromagnetic waves with an external constant electromagnetic field \cite{Schwinger1951}. The physical processes that would manifest are the light-light scattering, the photon splitting in an external electromagnetic field, pair creation for high field energies, and also pair creation in an external electromagnetic field. In the case of interaction of electromagnetic waves with an external magnetic field, essentially there is the light-light scattering with the external magnetic field where birefringence and dichroism effects would manifest depending on the energy of the incident electromagnetic wave. For a general discussion on the Euler-Heisenberg Lagrangian and its applications see Ref. \cite{Dunne:2004nc}.

In the last four decades, there have been several attempts to experimentally find the non-linear QED effects but so far such attempts have been quite elusive. One of the main reasons why it is difficult is because the magnitude of corrections to the Maxwell equations due to these non-linear effects is extremely small for typical laboratory magnetic field strengths and incident electromagnetic wave energies. The PVLAS experiment\cite{Zavattini:2005tm}, so far, has been one of the leading experiments trying to find and measure the vacuum birefringence in a constant magnetic field as predicted by Euler and Heinsenberg\cite{Heisenberg:1935qt} theory. The main methods to probe these non-linear QED effects that have been employed by the PVLAS collaboration are measurements of the rotation of the polarization plane of the incident electromagnetic wave and the induced ellipticity in the presence of a constant and transverse external magnetic field.

The Euler-Heisenberg theory and its applicability in physics go beyond laboratory searches of non-linear QED effects to the Maxwell equations and it can be applied in many cosmological and astrophysical situations as well, where interaction among electromagnetic fields occur. For example, in cosmology it is very well known that interaction of the CMB photons with a cosmic magnetic field would generate CMB circular polarization due to the application of the Euler-Heisenberg theory, see Ref. \cite{Ejlli:2016avx} for details. However, so far in all studies of the vacuum birefringence, the external magnetic field has a fixed direction with respect to the electromagnetic wave direction of propagation, and it is quite important to study the effects of vacuum birefringence for an arbitrary direction of the eternal magnetic field. This fact is mainly motivated because the vacuum birefringence in the cosmic magnetic field, which generates CMB circular polarization, most likely occurs in a cosmic magnetic field which direction is not known a priory and which can be arbitrary. This ignorance about the direction of the cosmic magnetic field must be properly accounted for by calculating explicitly the elements of the photon polarization tensor in a constant magnetic field of arbitrary direction. Besides, as I will show in this work, in the case when the external magnetic field has a longitudinal component with respect to the direction of propagation of the electromagnetic wave, a longitudinal electric field (longitudinal photon state) would appear in a magnetized vacuum. I show that the mixing of the longitudinal and transverse states of the electromagnetic radiation causes a decrease of the group velocity of electromagnetic waves in a magnetized vacuum. Given these two realistic situations that have never been studied in the literature and given their importance, in this work I study the non-linear QED effect to the Maxwell equation for a constant external magnetic field of arbitrary direction and study the implications of the longitudinal photon state to the wave propagation and velocity in a magnetized vacuum. Then I apply these findings to realistic cases of distance measurements in cosmology.

This work is organized as follows: In Sec. \ref{sec:2}, I find the constitutive equations of the electromagnetic field in a constant magnetic due to non linear QED effects. In Sec. \ref{sec:3}, I calculate the elements of the photon polarization tensor in a constant external magnetic field. In Sec. \ref{sec:4}, I calculate the group velocity of the electromagnetic waves in a magnetized vacuum. In Sec. \ref{sec:5}, I apply the results obtained to cosmological distance measures and then I conclude in Sec. \ref{sec:6}. In this work I use the metric with signature $\eta_{\mu\nu}=\text{diag}[1, -1, -1, -1]$ and work with the rationalized Lorentz-Heaviside natural units ($k_B=\hbar=c=\varepsilon_0=\mu_0=1$) with $e^2=4\pi \alpha$.

\section{Constitutive field equations in a magnetized vacuum}
\label{sec:2}

Consider the full expression of the renormalized Euler-Heisenberg Lagrangian density which is given by\cite{Heisenberg:1935qt}, \cite{Dunne:2004nc}
\begin{equation}\label{EH-lan}
\mathcal L_\text{EH} = -\frac{1}{8\pi^{2}}\int_{0}^{\infty}\exp\left(-m_e^{2}s\right)\left[(es)^{2}\mathcal G \frac{\operatorname{Re}\cosh\left(es\sqrt{2\left(\mathcal{F} + i\mathcal{G}\right)}\right)}{\operatorname{Im}\cosh\left(es\sqrt{2\left(\mathcal{F} + i\mathcal{G}\right)}\right)}-\frac{2}{3}(es)^{2}\mathcal{F} - 1\right]\frac{ds}{s^{3}},
\end{equation}
where $s$ is the variable of integration, $e$ is the electron charge and $\mathcal F, \mathcal G$ are the usual Lorentz invariant functions that are respectively defined as
\begin{equation}
\mathcal F=\frac{1}{4} F_{\mu\nu} F^{\mu\nu}=\frac{1}{2} (\bs B^2-\bs E^2), \quad \mathcal G= \frac{1}{4} F_{\mu\nu} F^{*\mu\nu}=-\bs E\cdot \bs B.
\end{equation}
The Euler-Heisenberg Lagrangian density \eqref{EH-lan} describes the one-loop correction\footnote{Calculation of the low energy Euler-Heisenberg Lagrangian for N photon amplitudes, two-loops correction and the analogy between one-loop photon-photon scattering with the photon-graviton scattering have been studied in Refs. \cite{Martin:2003gb}.} to the Maxwell equations in vacuum to all orders in perturbation theory. One of the most important aspects of expression \eqref{EH-lan} is that there are not present space-time derivatives of the electromagnetic field tensor $F_{\mu\nu}$ since it is valid only for constant or slowly varying spacetime electromagnetic fields with respect to the Compton wavelength. This essentially means that \eqref{EH-lan} is valid in the case when the energies of the interacting electromagnetic fields are smaller than the electron mass $\omega\ll m_e$ or that the spatial variation of the interacting electromagnetic fields are bigger than the Compton wavelength. The expression for \eqref{EH-lan} agrees with the results found in Ref. \cite{Karplus:1950zz} in the case of slowly varying spacetime electromagnetic fields and it is usually called the low energy correction Lagrangian density in the vacuum.

Let us now define for simplicity the critical field strength as $B_c=m_e^2/e$ and assume that the electromagnetic field strength is much smaller than the critical field strength, namely $\sqrt{\mathcal F}/B_c\ll 1$ and $\sqrt{\mathcal G}/B_c\ll 1$. In this case we can expand in series as 
\begin{equation*}
\cosh \left(es\sqrt{2\left(\mathcal{F} + i\mathcal{G}\right)}\right)= 1 + (es)^2 (\mathcal F+i\mathcal G) + \frac{(es)^4}{6} (\mathcal F+i\mathcal G)^2 +  \frac{(es)^6}{90} (\mathcal F+i\mathcal G)^3 +...
\end{equation*}
Then we have that 
\begin{eqnarray}\label{cosh-IR}
\text{Re} \cosh \left(es\sqrt{2\left(\mathcal{F} + i\mathcal{G}\right)}\right) &=& 1 + (es)^2 \mathcal F+ \frac{(es)^4}{6} \left(\mathcal F^2 -\mathcal G\right)^2 +  \frac{(es)^6}{90} \left(\mathcal F^3 -3 \mathcal F \mathcal G^2\right) +...\nonumber\\
\text{Im} \cosh \left(es\sqrt{2\left(\mathcal{F} + i\mathcal{G}\right)}\right) &=& (es)^2 \mathcal G+ \frac{(es)^4}{3} \mathcal F\mathcal G +  \frac{(es)^6}{90} \left(3\mathcal F^2 \mathcal G-\mathcal G^3\right) +...
\end{eqnarray}
Now we can use the results in \eqref{cosh-IR} and get the following result for 
\begin{equation}\label{cosh-1}
(es)^{2}\mathcal G \frac{\operatorname{Re}\cosh\left(es\sqrt{2\left(\mathcal{F} + i\mathcal{G}\right)}\right)}{\operatorname{Im}\cosh\left(es\sqrt{2\left(\mathcal{F} + i\mathcal{G}\right)}\right)} = 1+\frac{2 (es)^2}{3} \mathcal F -\frac{(es)^4}{45} \left(4 \mathcal F^2 +7 \mathcal G^2\right) +...
\end{equation}
Now by using \eqref{cosh-1} in expression \eqref{EH-lan}, we get 
\begin{equation}\label{EH-lan-1}
\mathcal L_\text{EH} = -\frac{1}{8\pi^{2}}\int_{0}^{\infty}\exp\left(-m_e^{2}s\right)\left[(es)^{2}\mathcal G \frac{\operatorname{Re}\cosh\left(es\sqrt{2\left(\mathcal{F} + i\mathcal{G}\right)}\right)}{\operatorname{Im}\cosh\left(es\sqrt{2\left(\mathcal{F} + i\mathcal{G}\right)}\right)}-\frac{2}{3}(es)^{2}\mathcal{F} - 1\right]\frac{ds}{s^{3}} = \frac{2\alpha^2}{45 m_e^4} \left(4 \mathcal F^2 +7 \mathcal G^2\right) +...
\end{equation}
From relation \eqref{EH-lan-1} we obtain the expression for the weak and slowly varying (or constant) interacting electromagnetic fields Lagrangian density 
\begin{equation}\label{EH-lan-2}
\mathcal L_\text{EH}(\bs E, \bs B) = \frac{2\alpha^2}{45 m_e^4} \left[4 (\bs E^2-\bs B^2)^2 +7 (\bs E\cdot \bs B)^2\right] + ...
\end{equation}
 
The Lagrangian density \eqref{EH-lan-2} describes non linear effects in QED as it can be seen from its structure. These non-linear effects can be studied by finding the corresponding electric displacement $\bs D$ and magnetic magnetization $\bs H$ vectors from the Lagrangian \eqref{EH-lan-2}. In general, in the case of propagation of electromagnetic waves in media, the total Lagrangian density in absence of a net free charge ($\rho_\text{free}=0$) and free currents ($\bs J_\text{free}=0$), is given by
\begin{equation}\label{int-lan}
\mathcal L= -\frac{1}{4} F_{\mu\nu}F^{\mu\nu} - \frac{1}{2} F_{\mu\nu} M^{\mu\nu},
\end{equation}
where $M^{\nu\nu}$ is the covariant magnetization-polarization tensor in media which is given by 
\begin{equation}\label{D-matrix}
M^{\mu\nu}=
\begin{pmatrix}
0 & - P_x & - P_y &  P_z\\
 P_x & 0 & - M_z &  M_y \\
 P_y &  M_z & 0 & - M_x \\
 P_z & - M_y & M_x & 0
\end{pmatrix}.
\end{equation}
It can be shown that by substituting the expression for $M^{\mu\nu}$ in \eqref{D-matrix} into \eqref{int-lan}, the latter becomes
\begin{equation}\label{E-D}
\mathcal L(\bs E, \bs B) =\frac{1}{2}\left( \bs E^2 - \bs B^2 \right) + \bs E \bs P + \bs B \bs M,
\end{equation}
where $\bs P$ is the polarization density vector in media and $\bs M$ is the magnetization vector in the same media. The expression for the Lagrangian density in \eqref{E-D} is the most general one for an electromagnetic field propagating in a dispersive medium. We may observe that the equation of motions for the electric $\bs E$ and magnetic $\bs B$ fields are given by
\begin{equation}\label{H-D-eq}
\frac{\partial\mathcal L(\bs E, \bs B)}{\partial\bs E}=\bs E+ \bs P \equiv \bs D, \quad -\frac{\partial\mathcal L(\bs E, \bs B)}{\partial\bs B} = \bs B - \bs M \equiv \bs H,
\end{equation}
where $\bs D$ and $\bs H$ are respectively the electric induction and magnetic intensity vectors.

By writing the total Lagrangian density as $\mathcal L(\bs E, \bs B)=\mathcal L_0(\bs E, \bs B) + \mathcal L_\text{EH}(\bs E, \bs B)$ where $\mathcal L_0$ is the free electromagnetic field Lagrangian density, we get from \eqref{H-D-eq} the following expressions
\begin{eqnarray}\label{H-D-eq-1}
\bs D &=& \bs E + \frac{2 \alpha^2}{45 m_e^4} \left[ 4\, \bs E\, (\bs E^2 -\bs B^2) +14\, \bs B\, (\bs E\cdot \bs B) \right], \nonumber\\
\bs H &=& \bs B + \frac{2 \alpha^2}{45 m_e^4} \left[ 4\, \bs B\, (\bs E^2 -\bs B^2) -14\, \bs E\, (\bs E\cdot \bs B) \right].
\end{eqnarray}
The expression in \eqref{H-D-eq-1} is quite general and gives the so-called constitutive equations of the electromagnetic field in a magnetized vacuum. Now suppose we write the total electric and magnetic fields as $\bs E= \bs E_\gamma +\bar{\bs E}$ and $\bs B = \bs B_\gamma + \bar{\bs B}$ where $\bs E_\gamma$ and $\bs B_\gamma$ are respectively the electric and magnetic field vectors of the incident propagating electromagnetic field. On the other hand, $\bar{\bs E}$ and $\bar{\bs B}$ are respectively the electric and magnetic field vectors of the background electromagnetic field. We may observe from \eqref{H-D-eq-1} that in the case of plane electromagnetic waves there are not corrections to the Maxwell equations due to vacuum polarization, namely the interaction of plane electromagnetic waves does not produce non-linear effects because of the invariants $\bs E^2 -\bs B^2=0$ and $\bs E\cdot \bs B=0$. This is an important fact to keep always in mind when we deal with non-linear QED corrections to the vacuum Maxwell equations.

Now consider the case when only an external magnetic field is present and no external electric field $\bar{\bs E}=0$. Besides, we consider the case when the external magnetic field is constant in space and time, namely a uniform field. In this case, we get
\begin{eqnarray}\label{H-D-eq-2}
\bs D_\gamma &=& \bs E_\gamma + \frac{2 \alpha^2}{45 m_e^4} \left[ 4\, \bs E_\gamma\, (\bs E_\gamma^2 -\bs B^2) +14\, \bs B\, (\bs E_\gamma {\bs B)} \right], \nonumber\\
\bs H &=& \bs B + \frac{2 \alpha^2}{45 m_e^4} \left[ 4\, \bs B\, (\bs E_\gamma^2 -\bs B^2) -14\, \bs E_\gamma\, (\bs E_\gamma {\bs B}) \right].
\end{eqnarray}
%where we considered the fact that for the incident electromagnetic wave $\bs B_\gamma \bs E_\gamma=0$.
If in \eqref{H-D-eq-2} we consider only the linear terms in the weak (with respect to the external magnetic field, $|\bs E_\gamma|, |\bs B_\gamma| \ll |\bar{\bs B}|$) incident electromagnetic wave and neglect higher-order terms, we get the following linearized equations of motion
\begin{eqnarray}\label{H-D-eq-3}
\bs D_\gamma & \simeq & \left[1 - \frac{8 \alpha^2}{45 m_e^4} {\bar B}^2 \right] \bs E_\gamma + \frac{28 \alpha^2}{45 m_e^4} \, \bar{\bs B}\, (\bs E_\gamma \bar{\bs B)}, \nonumber\\
\bs H_\gamma + \bar{\bs H} & \simeq &  \left[1  - \frac{8 \alpha^2}{45 m_e^4}\,  {\bar B}^2  \right]\left(\bar{\bs B} + \bs B_\gamma \right)-\frac{16 \alpha^2}{45 m_e^4}\, \bar{\bs B} (\bs B_\gamma \bar{\bs B}).
\end{eqnarray}

\section{Calculations of the elements of the photon polarization tensor}
\label{sec:3}

At this point let us recall that the Maxwell equations in matter are given by
\begin{equation}\label{Max}
\nabla\cdot \bs D = 0, \quad \nabla\cdot \bs  B = 0, \quad \nabla\times \bs H = \partial_t \bs D, \quad \nabla\times \bs E = - \partial_t\bs B.
\end{equation}
Using the fact that the external magnetic field ($\bar{\bs B}$) is assumed to be constant and no external electric field ($\bar{\bs E} = 0 $) where $\bs D = \bs D_\gamma$ and $\bs H = \bs H_\gamma + \bar{\bs H}$, we get the following set of equations for the propagating fields\footnote{Generally speaking by using \eqref{H-D-eq-3} and Maxwell's equations \eqref{Max} there are two sets of equations, one for the background fields and another one for the propagating fields. Here we are interested in the equations of motion for the propagating fields only. Clearly, if the external fields are constant and uniform as have been assumed so far in this work, we can completely neglect their spacetime derivatives to the spacetime derivatives of the propagating fields.}
\begin{equation}\label{Max-1}
\nabla\cdot \bs D_\gamma = 0, \quad \nabla\cdot \bs  B_\gamma = 0, \quad \nabla\times \bs H_\gamma = \partial_t \bs D_\gamma, \quad \nabla\times \bs E_\gamma = - \partial_t\bs B_\gamma.
\end{equation}
Maxwell equations in \eqref{Max-1} can be put in a more elegant way if we write  $D_\gamma^i=\varepsilon_{ij} E_\gamma^j$, $H_\gamma^i =\mu_{ij}^{-1} B_\gamma^j$ where $\varepsilon_{ij}$ and $\mu_{ij}$ are respectively the electric permittivity and magnetic permeability tensors due to the interaction of the electromagnetic wave with the external magnetic field. Their expressions are respectively given by
\begin{equation}\label{e-mu-tens}
\varepsilon_{ij} = \delta_{ij} \left[1 - \frac{8 \alpha^2}{45 m_e^4} {\bar B}^2 \right] + \frac{28 \alpha^2}{45 m_e^4}\bar B_i \bar B_j, \quad \mu_{ij}^{-1} = \delta_{ij} \left[1 - \frac{8 \alpha^2}{45 m_e^4} {\bar B}^2 \right] - \frac{16 \alpha^2}{45 m_e^4}\bar B_i \bar B_j .
\end{equation}
By going into momentum space with Fourier transform, we get from Maxwell equations in \eqref{Max-1} and from \eqref{e-mu-tens} the following sets of algebraic equations
\begin{eqnarray}\label{e-mu-eqs}
ik^l\varepsilon_{lj} E_\gamma^j(\bs k, \omega)&=&  - k^l\varepsilon_{lj} \omega A_\gamma^j(\bs k, \omega) =0,\\
\left[\varepsilon_{ij} + \epsilon_{ikl} \epsilon_{mrj} n_k n_r \mu_{lm}^{-1} \right] E_\gamma^{j}(\bs k, \omega) &=&  \left[\varepsilon_{ij} + \epsilon_{ikl} \epsilon_{mrj} n_k n_r \mu_{lm}^{-1} \right] i \omega A_\gamma^{j}(\bs k, \omega)=0,
\end{eqnarray}
where $\epsilon_{ikl}$ is the three dimensional Levi-Civita symbol, $n_{k, r}$ are the components of the refraction index which is defined as $\bs n = \bs k/\omega$ and $A_\gamma^j$ are the spatial components of the vector potential $\bs A$. Here we are working in the temporal gauge where by choice $A^0=0$.

It is quite common to work with the electric $\chi_{ij}^{(e)}$ and magnetic $\chi_{ij}^{(m)}$ susceptibilities tensors instead of $\varepsilon_{ij}$ and $\mu_{ij}$. The latter tensors can be expressed as $\varepsilon_{ij}=\delta_{ij} + \chi_{ij}^{(e)}$ and $\mu_{ij}= \delta_{ij} + \chi_{ij}^{(m)}$, where from \eqref{e-mu-tens} we obtain the following expressions
\begin{equation}
\chi_{ij}^{(e)} = - \frac{8 \alpha^2}{45 m_e^4} {\bar B}^2 \delta_{ij} + \frac{28 \alpha^2}{45 m_e^4}\bar B_i \bar B_j, \quad \chi_{ij}^{(m)} \simeq   \frac{8 \alpha^2}{45 m_e^4} {\bar B}^2 \delta_{ij} + \frac{16 \alpha^2}{45 m_e^4}\bar B_i \bar B_j ,
\end{equation}
where we used $\mu_{ij}^{-1} \simeq \delta_{ij} - \chi_{ij}^{(m)}$ to first order for $\bar B \ll B_c$.

Let us recall that the covariant version of Maxwell equations are $\partial_\mu F^{\mu\nu}=J^\nu$ and $\partial_\mu F^{*\mu\nu}=0$. In momentum space the first set of Maxwell equations ($\partial_\mu F^{\mu\nu}=J^\nu$) in terms of the vector potential $A^\mu$ are 
\begin{equation}\label{cov-Max}
\left[ k^2 g^{\mu\nu} - k^\mu k^\nu \right] A_\nu(k) = -J^\mu (k),
\end{equation}
where $k^\mu= (\omega, \bs k)$ and $k^2=\omega^2 - \bs k^2$. In absence of free charges and currents, we have that only induced currents are present where $J_\text{ind}^{\mu}(k)=\Pi^{\mu\nu}(k) A_\nu(k)$ where $\Pi^{\mu\nu}$ is the photon polarization tensor. In the temporal gauge $A^0=0$, we have from \eqref{cov-Max}
\begin{eqnarray}\label{e-mu-cov}
[k_0 k_j - \Pi_{0j}(\bs k, \omega) ] A_\gamma^j(\bs k, \omega) & = & 0,\qquad (\mu=0), \nonumber \\
\left [(\omega^2 - \bs k^2 )\,\delta_{ij} + k_i k_j - \Pi_{ij}(\bs k, \omega) \right] A_\gamma^j(\bs k, \omega) & = &0, \qquad (\mu=i).
\end{eqnarray}
One important property of the photon polarization tensor is that at the linear order correction, it must satisfy the charge continuity equation $k_\mu J^\mu=0$ and gauge invariance. It can be shown that under such conditions, we have the following conditions
\begin{equation}\label{CC-eq}
k_\mu\Pi^{\mu\nu}(\bs k, \omega)=0, \qquad k_\nu\Pi^{\mu\nu}(\bs k, \omega)=0.
\end{equation}
Now by writing the equations in \eqref{e-mu-eqs} in terms of the electric and magnetic susceptibility tensors and then comparing them with the scalar and vectorial equations in \eqref{e-mu-cov}, we get after few steps the following relations
\begin{eqnarray}\label{Pi-form}
\Pi_{0j}(\bs k, \omega)  &= & -\omega^2 n^i\chi_{ij}^{(e)},\nonumber\\
\Pi_{ij}(\bs k, \omega) &= &-\omega^2\left[ \chi_{ij}^{(e)} - \epsilon_{ikl}\epsilon_{mrj} n_k n_r \chi_{lm}^{(m)} \right].
\end{eqnarray}
We can observe that expressions of the photon polarization elements in \eqref{Pi-form} identically satisfy the charge continuity and gauge invariance conditions in \eqref{CC-eq}.

In order to explicitly calculate the elements of the photon polarization tensor, we assume without loss of generality that $\bs k = (0, 0, |\bs k|)$, namely an electromagnetic wave propagating along the $z$ axis in a cartesian coordinate system. In addition, we assume that the external magnetic field vector has an arbitrary direction with respect to the electromagnetic wave direction of propagation, $\bar{\bs B} = (\bar B_x, \bar B_y, \bar B_z)$. By using expression \eqref{Pi-form}, we get the following matrix elements for the spatial components of the photon polarization tensor 
\begin{equation}\label{El-Pi}
\begin{gathered}
\Pi_{xx} (k, \omega) =  - \frac{\alpha^2}{45 m_e^4} \left[ 8 \bar B^2\left( k^2 - \omega^2\right) + 28 \bar B_x^2\omega^2+16\bar B_y^2k^2  \right], \, \Pi_{xy} (k, \omega) = - \frac{4 \alpha^2}{45 m_e^4} \left[ 7\omega^2-4k^2 \right] \bar B_x\bar B_y, \, \Pi_{xz} (k, \omega)= - \frac{28\alpha^2\omega^2}{45 m_e^4} \bar B_x\bar B_z, \\
\Pi_{yx} (k, \omega)= \Pi_{xy}(k, \omega), \quad \Pi_{yy}(k, \omega)= - \frac{\alpha^2}{45 m_e^4} \left[ 8 \bar B^2\left( k^2 - \omega^2\right) + 28 \bar B_y^2\omega^2+16\bar B_x^2 k^2 \right], \quad \Pi_{yz}(k, \omega)= - 28 \frac{\alpha^2\omega^2}{45 m_e^4} \bar B_y\bar B_z, \\  \Pi_{zx}(k, \omega) =  \Pi_{xz} (k, \omega), \quad
\Pi_{zy} (k, \omega)= \Pi_{yz}(k, \omega), \quad \Pi_{zz} (k, \omega)= - \frac{\alpha^2\omega^2}{45 m_e^4} \left[ -8 \bar B^2 + 28\bar B_z^2 \right],
\end{gathered}
\end{equation}
where $k=|\bs k|$. We may observe that the elements of the photon polarization tensor in \eqref{El-Pi} satisfy the condition $\Pi_{ij}(k, \omega)= \Pi_{ji}^*(k, \omega)$. This condition it can be shown to imply that the total number of photons is a conserved quantity.

\section{Longitudinal photon state and group velocity}
\label{sec:4}

With the aid of equations in \eqref{e-mu-cov} we can now derive the dispersion relations for the electromagnetic waves propagating in a magnetized medium. Assume as in the previous section that $\bs k = (0, 0, |\bs k|)$, namely propagation along the $z$ axis and $\bar{\bs B} = (\bar B_x, \bar B_y, \bar B_z)$. In this case we get from \eqref{e-mu-cov}  the following system of linear equations
\begin{equation}\label{system-al-1}
\begin{gathered}
\omega k A_\gamma^z(k, \omega) - \Pi_{0x}(k, \omega) A_\gamma^x(k, \omega) - \Pi_{0y}(k, \omega) A_\gamma^y(k, \omega) - \Pi_{0z}(k, \omega) A_\gamma^z(k, \omega) =0, \quad (\mu=0)\\
(\omega^2 - k^2) A_\gamma^x(k, \omega)  - \Pi_{xx}(k, \omega) A_\gamma^x(k, \omega) - \Pi_{xy}(k, \omega) A_\gamma^y(k, \omega)- \Pi_{xz}(k, \omega) A_\gamma^z(k, \omega ) = 0, \quad (\mu=1)\\
(\omega^2 - k^2) A_\gamma^y(k, \omega)  - \Pi_{yx}(k, \omega) A_\gamma^x(k, \omega) - \Pi_{yy}(k, \omega) A_\gamma^y(k, \omega)- \Pi_{yz}(k, \omega) A_\gamma^z(k, \omega) = 0, \quad (\mu=2)\\
\left[ \omega^2 - \Pi_{zz}(k, \omega) \right] A_\gamma^z (k, \omega) - \Pi_{zx}(k, \omega) A_\gamma^x(k, \omega) - \Pi_{zy}(k, \omega) A_\gamma^y(k, \omega) = 0,  \quad (\mu=3),
\end{gathered}
\end{equation}
where $i=x, y, z$. One important thing about the system of equations in \eqref{system-al-1} is that the first equation for $\mu=0$ is identical to the fourth equation for $\mu=3$. This can be seen by using the fact that from \eqref{CC-eq} we have the property that $\Pi_{0j}= k^i\Pi_{ij}/k_0$ where the sum over repeated indexes is intended. 

With this fact in mind, we have to solve only the system of equations for the $\mu=1, 2, 3$ which can be written as $C_{ij}(k, \omega) A^i(k, \omega)=0$, where the coefficient matrix is given by
\begin{equation}\label{matrix-C}
C_{ij}(k, \omega)=\begin{pmatrix}
\omega^2 - k^2 - \Pi_{xx}(k, \omega) & - \Pi_{xy}(k, \omega)  & - \Pi_{xz}(k, \omega)   \\
- \Pi_{xy}(k, \omega) & \omega^2 - k^2 - \Pi_{yy}(k, \omega)& -\Pi_{yz}(\omega) \\
-\Pi_{xz}(k, \omega)  & -\Pi_{yz}(\omega)  & \omega^2 - \Pi_{zz}(k, \omega)  \\
   \end{pmatrix},      
\end{equation}
where we used the fact that the elements of the photon polarization tensor are real and that $\Pi_{ij}(k, \omega) = \Pi_{ji}(k, \omega)$. The system of linear equations $C_{ij}(k, \omega) A_\gamma^i(k, \omega)=0$ has solutions if and only if Det[$C_{ij}(k, \omega)]=0$. The determinant of the matrix $C_{ij}$ in \eqref{matrix-C} is given by
\begin{equation}\label{Det-1}
\begin{gathered}
\text{Det} [C_{ij}(k, \omega)] =  \left[ \omega^2 - k^2 - \Pi_{xx}(k, \omega)\right]   \left( \left[ \omega^2 - \Pi_{zz}(k, \omega) \right]  \left[ \omega^2 - k^2 - \Pi_{yy}(k, \omega) \right] - [\Pi_{yz}(k, \omega)]^2 \right) \\ - \Pi_{xy}(k, \omega) \left[ \Pi_{xy}(k, \omega) \left( \omega^2 - \Pi_{zz}(k, \omega) \right) +  \Pi_{yz}(k, \omega)  \Pi_{xz}(k, \omega) \right]    - \Pi_{xz}(k, \omega) \left[ \Pi_{xz}(k, \omega) \left( \omega^2 - k^2 - \Pi_{yy}(k, \omega) \right) +  \Pi_{xy}(k, \omega)  \Pi_{yz}(k, \omega) \right].
\end{gathered}
\end{equation}
By using the expressions of the elements of the photon polarization tensor in \eqref{El-Pi} in \eqref{Det-1} and doing lengthy algebraic operations and manipulations, the condition Det[$C_{ij}(k, \omega)]=0$, reduces to the following equation
\begin{equation}\label{polinom}
\omega^2 \left( 1- 8\rho \bar B^2 \right) \left[ \omega^2 - k^2 + 4\rho \left( (\bar B_x^2 + \bar B_y^2)  (5\omega^2 + 2 k^2)  + 5 \bar B_z^2 (\omega^2 - k^2) \right) \right] \left[ \omega^2 - k^2 + 8\rho \left( (\bar B_x^2 + \bar B_y^2)  (3k^2 - \omega^2)  - \bar B_z^2 (\omega^2 - k^2) \right) \right] = 0,
\end{equation}
where we have defined $\rho\equiv \alpha^2/(45 m_e^4)=\alpha/(180\pi) (1/B_c)^2$.

Equation \eqref{polinom} depends on several variables and as it is common in the literature we look for its solution in the case when $\omega$ is considered to be the independent variable which is usually a function of $k$, namely $\omega=\omega(k)$. In this case Eq. \eqref{polinom} is a polynomial equation of power six in $\omega$ and in principle has six roots. First of all we exclude the trivial solution $\omega=0$ that has double multiplicity because we are working under the hypothesis of non static fields, namely $\omega\neq 0$. Another case when \eqref{polinom} would be zero is when $\rho \bar B^2 =1/8$ for any $\omega$ and $k$. However, even this case is to be excluded because we are working under the hypothesis that $\bar B/B_c\ll 1$, see Sec. \ref{sec:2} for details, so $\rho \bar B^2 =1/8$ is never satisfied. The only possibility for equation \eqref{polinom} to be zero if either one or both terms within square parenthesis are zero. By solving each equation within square parenthesis, we find the following solutions
\begin{equation}\label{omega-sol}
\omega_1 \equiv \omega = \pm k \left[  \frac{1 - 8\rho (\bar B_x^2 + \bar B_y^2) +20 \rho\bar B_z^2}{1 + 20 \rho \bar B^2}  \right]^{1/2}, \qquad \omega_2 \equiv \omega = \pm k \left[  \frac{1 - 24\rho (\bar B_x^2 + \bar B_y^2) -8 \rho\bar B_z^2}{1 - 8 \rho \bar B^2}  \right]^{1/2}.
\end{equation}
Given the fact that $\rho \bar B^2 \ll 1$, we essentially have that the square root of each term in the numerator and denominator in \eqref{omega-sol} is a positive and real quantity and consequently $\omega(k)$ is a real quantity. This reflects the fact that we are working under the condition $\omega\ll m_e$ where pair creation in the magnetic field is not allowed and consequently the photon number is a conserved quantity. This translates also into the condition that all elements of $\Pi_{ij}(k, \omega)$ are real. By considering for simplicity the positive solutions in \eqref{omega-sol} and expanding in series the denominators for $\rho \bar B^2\ll 1$, we get the following expressions for the the phase velocity ($v_\text{p} = \omega(k)/k$) and group velocity ($v_\text{g} = \partial_k \omega(k)$) of the electromagnetic waves propagating in a magnetized vacuum 
\begin{equation}\label{gp-velo}
v_{1, \text{p}} = v_{1, \text{g}} \simeq 1 - 14 \rho (\bar B_x^2 + \bar B_y^2), \qquad v_{2, \text{p}} = v_{2, \text{g}} \simeq 1 - 8 \rho (\bar B_x^2 + \bar B_y^2).
\end{equation}
The expressions of the velocities in \eqref{gp-velo} tell us that for each mode of propagation, the group and phase velocities are equal, they do not depend on the longitudinal part of the external magnetic field $\bar B_z$ at first-order expansion in the quantity $\rho \bar B^2\ll 1$ and they are smaller than $c=1$.

With the solutions given in \eqref{omega-sol}, we can also entirely write the expressions of the elements of the photon polarization tensor in \eqref{El-Pi} as a function of $\omega^2$ or $k^2$. If we choose to write them as a function of $\omega^2$, we can use both expressions in \eqref{omega-sol} and sum them with each other and after several manipulations we get $k^2 \simeq [1 + 12\rho (\bar B_x^2 + \bar B_y^2)] \omega^2$ where terms of the second order in the quantities $\rho \bar B^2\ll 1, \rho \bar B_z^2\ll 1, \rho (\bar B_x^2 + \bar B_y^2)\ll 1$ have been neglected. With  $k^2 \simeq [1 + 12\rho (\bar B_x^2 + \bar B_y^2)] \omega^2$ and replacing it in \eqref{El-Pi}, we get
\begin{equation}\label{El-Pi-1}
\begin{gathered}
\Pi_{xx} (\omega) \simeq  - \frac{\alpha^2\omega^2}{45 m_e^4} \left[ 28 \bar B_x^2+16\bar B_y^2  \right], \, \Pi_{xy} (\omega) \simeq - \frac{12 \alpha^2\omega^2}{45 m_e^4}\bar B_x\bar B_y, \, \Pi_{xz} (\omega)= - \frac{28\alpha^2\omega^2}{45 m_e^4} \bar B_x\bar B_z, \\
\Pi_{yx} (\omega)= \Pi_{xy}(\omega), \quad \Pi_{yy}(\omega) \simeq - \frac{\alpha^2\omega^2}{45 m_e^4} \left[ 28 \bar B_y^2+16\bar B_x^2 \right], \quad \Pi_{yz}(\omega)= - 28 \frac{\alpha^2\omega^2}{45 m_e^4} \bar B_y\bar B_z, \\  \Pi_{zx}(\omega) =  \Pi_{xz} (\omega), \quad
\Pi_{zy} (\omega)= \Pi_{yz}(\omega), \quad \Pi_{zz} (\omega)= - \frac{\alpha^2\omega^2}{45 m_e^4} \left[ -8 \bar B^2 + 28\bar B_z^2 \right],
\end{gathered}
\end{equation}
where clearly $\omega$ is explicitly a function of $k$, $\omega=\omega(k)$. So essentially, one can get expressions in \eqref{El-Pi-1} by simply replacing $k^2\simeq \omega^2$ in expressions in \eqref{El-Pi} and neglect smaller order corrective terms.

Another interesting fact is that in a magnetized vacuum it appears also a longitudinal state of the electromagnetic radiation. Indeed, from \eqref{system-al-1} we have that the longitudinal electric field is given by
\begin{equation}\label{long-state}
\left[ \omega^2 - \Pi_{zz}(k, \omega) \right] E_\gamma^z (k, \omega) = \Pi_{zx}(k, \omega) E_\gamma^x(k, \omega) + \Pi_{zy}(k, \omega) E_\gamma^y(k, \omega),
\end{equation}
where we used $E_\gamma^j(k, \omega)=i\omega A_\gamma^j(k, \omega)$ in the temporal gauge $A^0=0$. There is no magnetic field state associated to the longitudinal electric field $E_\gamma^z$ because of the equation $\bs k_\gamma \cdot \bs B_\gamma=0$ which implies $B_\gamma^z=0$ for $\bs k_\gamma$ along $z$ axis. Equation \eqref{long-state} indicates that the longitudinal electric field is a linear combination of transverse states of the electromagnetic wave and a net separation between the transverse states and the longitudinal state is no longer possible. All states are mixed with each other as far as a longitudinal component of the external magnetic field is present. To make this statement more clear, suppose that $\bar B_z=0$. Then from \eqref{El-Pi} we have that $\Pi_{zx, zy}=0$ and equation \eqref{long-state} reduces to $\left[ \omega^2 - \Pi_{zz}(k, \omega) \right] E^z (k, \omega) =0$. There are two possibilities for the latter equation to be satisfied, either $\omega^2 - \Pi_{zz}(k, \omega)=0$ or $E^z(k, \omega)=0$. By using the expression of $\Pi_{zz}$ in \eqref{El-Pi} we get that $\omega^2 - \Pi_{zz}(k, \omega)= \omega^2 -8 \omega^2\rho\bar B^2=0$. This equation has solution if $\omega=0$ or $\rho\bar B^2=1/8$. However, as we already have discussed above, both these solutions are excluded, so, the only possibility for  $\left[ \omega^2 - \Pi_{zz}(k, \omega) \right] E^z (k, \omega) =0$ is that $E^z(k, \omega)=0$ for $\bar B_z=0$. This analysis implies that a longitudinal electric field in a magnetized vacuum is excited only when the external magnetic field has a longitudinal component along the direction of propagation of the electromagnetic wave. This is very similar to the case of mixing of electromagnetic waves with a pseudoscalar field in vacuum where the longitudinal state is excited only when the external magnetic field has a longitudinal component\cite{Ejlli:2020ifc}.

\section{Application to cosmological distances}
\label{sec:5}

The results that we have found in the previous section about the group and phase velocities can have important applications to the case of distance measures in cosmology and the Cosmic Microwave Background (CMB). In general, it is always assumed that any form of electromagnetic radiation including the CMB, propagates through cosmological distances with constant velocity $c=1$, and based on this assumption many cosmological predictions are made. However, if cosmological magnetic fields existed in the early universe, then the velocity of light would not be constant anymore because of nonlinear correction of the velocity of the  electromagnetic waves as discussed in Sec. \ref{sec:4}. To investigate the change in the group velocity of the electromagnetic radiation, let us assume that cosmological magnetic fields exist in our universe and that the field amplitude changes in time as the universe expands, namely $\bar B=\bar B(\bs x, t)$ where $t$ is the cosmological time and $\bs x$ is the physical  or proper distance. Besides we also assume that during the cosmological expansion, the magnetic field flux is a conserved quantity which implies that $\bar B(\bs x, t) = \bar B_0(\bs x) [a(t_0)/a(t)]^2$ where $\bar B_0(\bs x)$ is the amplitude of the cosmic magnetic field at present time $t=t_0$ at the position $\bs x$ and $a(t)$ is the cosmological scale factor.

It is important to recall that all results in this work have been derived under the condition of a constant or at maximum for a  slowly varying external electromagnetic field since in the full Euler-Heisenberg Lagrangian density do not appear spacetime derivatives of the electromagnetic field tensor $\bar F_{\mu\nu}$, see Sec. \ref{sec:2} for details and Ref.\cite{Heisenberg:1935qt}.  
Consequently, our results found in the previous sections in the case of a constant magnetic field would also apply to the case when $|\partial_\mu F_{\sigma\rho}| \ll m_e \,|F_{\sigma\rho}|$. The latter condition must be satisfied for all interacting fields, namely for the propagating field that we are interested in and for the background field. In the case of the propagating electromagnetic field, the condition $|\partial_\mu F_{\sigma\rho}| \ll m_e \,|F_{\sigma\rho}|$ would imply in Fourier space that the electric and magnetic fields must have energies $\omega\ll m_e$ and wave-vectors $|\bs k| \ll m_e$ or $\lambda \gg 2\pi m_e^{-1}$ with $\lambda$ being the propagating electromagnetic wave wavelength. In an expanding universe both conditions $\omega\ll m_e$  and  $\lambda \gg 2\pi m_e^{-1}$ must be valid in the redshift\footnote{The redshift $z$ is defined as $a(t_0)/a(t) \equiv 1+z$ and is an increasing quantity as we go backward in time.} interval of interest $z\in [0, z_\text{max}]$ and both $\omega$ and $\lambda$ change with the redshift as well. If we write $\omega(z)=\omega_0(1+z)$, then the condition $\omega\ll m_e$ would be valid as far as $\omega_0\ll m_e/(1+z_\text{max})=7.75\times 10^{11}/(1+z_\text{max})$ GHz where $\omega_0$ is the present day value of the propagating electromagnetic wave energy and $z_\text{max}$ is the maximum value of the redshift that we are interested in. On the other hand by writing $\lambda = \lambda_0/(1+z)$, the condition $\lambda \gg 2\pi m_e^{-1}$, translates into  $\lambda_0 \gg 2\pi m_e^{-1}(1+z_\text{max})= 2.43\times 10^{-10}(1+z_\text{max})$ cm.
In an expanding universe and in the presence of an external magnetic field only, the condition $|\partial_\mu \bar F_{\sigma\rho}| \ll m_e \,|\bar F_{\sigma\rho}|$ implies that $|\partial_t \bar B^i(\bs x, t)|\ll m_e |\bar B^i(\bs x, t)|$ and $|\partial_j \bar B^i(\bs x, t)|\ll m_e |\bar B^i(\bs x, t)|$. We have that
\begin{equation}
\partial_t \bar B^i(\bs x, t) = - 2 H(t) \bar B^i(\bs x, t),
\end{equation}
where we used $\bar B(\bs x, t) = \bar B_0(\bs x) [a(t_0)/a(t)]^2$. So, the condition $|\partial_t \bar B^i(\bs x, t)|\ll m_e |\bar B^i(\bs x, t)|$ translates into a condition for the Hubble time $H^{-1}(z_\text{max})\gg 2/m_e \simeq 2.58 \times 10^{-21}$ s, while the condition $|\partial_j \bar B^i(\bs x, t)|\ll m_e |\bar B^i(\bs x, t)|$ in momentum space translates into $l_{\bar B}(z_\text{max})\gg 2.43\times 10^{-10}(1+z_\text{max})$ cm, where $\bar B^i$ is the $i$-th component of $\bs  \bar B$ and $l_{\bar B}$ is the variation scale in space of external magnetic field corresponding to the physical wave-vector $\bar{\bs k}$.  Here $H(t) = \dot a(t)/a(t)$ is the Hubble parameter as a function of the scale factor $a(t)$.

After having established the conditions of validity of our results, consider for example the case of the light travelled distance in an expanding universe, which expression is given by
\begin{equation}\label{light-t-distance}
d_c(z)=\int_{0}^{z}\,\frac{c\, dz^{\prime}}{H_0(1+z^\prime)\sqrt{\Omega_\Lambda+\Omega_\text{M}(1+z^\prime)^3+\Omega_\text{R}(1+z^\prime)^4}},
\end{equation}
where $\Omega_\Lambda\simeq 0.68$ is the present epoch density parameter of the vacuum energy, $\Omega_\text{M}\simeq 0.31$ is the present epoch density parameter of the nonrelativistic matter, $\Omega_\text{R}\ll 1$ is the present epoch density parameter of the relativistic matter that essentially includes relativistic photons and neutrinos and $H_0=H(t_0)$ is the present day Hubble parameter. Here we are assuming a universe with zero spatial curvature, namely $\Omega_\kappa=0$.  In expression \eqref{light-t-distance} we have restored the velocity of light in vacuum $c=1$ as it will useful in what follows and the subscript $c$ in $d_c(z)$ indicates the light travelled distance when the light propagates with velocity. In the case when the group velocity is given by \eqref{gp-velo}, we have that the light travelled distance is given by
\begin{equation}\label{light-t-distance}
d_{\bar v}(z)=\int_{0}^{z}\,\frac{c\,\bar v(z^\prime)\, dz^{\prime}}{H_0(1+z^\prime)\sqrt{\Omega_\Lambda+\Omega_\text{M}(1+z^\prime)^3+\Omega_\text{R}(1+z^\prime)^4}},
\end{equation}
where $\bar v(z)$ is the average value of the group velocity over the solid angle $\Omega$ and polarization states of the electromagnetic radiation. The dependence of the group velocity on $z$ is due to the fact that it implicitly depends on $z$ through amplitude of the magnetic field $\bar B(z)$. The difference on the light travelled distance is given by
\begin{equation}\label{delta-d}
\Delta d^\text{LT}(z) \equiv d_c(z) - d_{\bar v} (z) = c H_0^{-1} \int_{0}^{z}\,\frac{[1 - \bar v(z^\prime)]\, dz^{\prime}}{(1+z^\prime)\sqrt{\Omega_\Lambda+\Omega_\text{M}(1+z^\prime)^3+\Omega_\text{R}(1+z^\prime)^4}}.
\end{equation}

Now we need to calculate the average group velocity $\bar v(z)$ that appear in \eqref{light-t-distance} where $\bar v(z)$ depends on $z$ only through the magnetic field amplitude $\bar B_{x, y}^2$. 
The average value of the group velocity over the two polarization modes and over the volume is given by
\begin{equation}\label{av-velo}
\bar v(z) = \langle (v_{1, g} + v_{2, g})/2 \rangle = 1-11\rho (1+z)^4  \bar B_{0, T}^2,
\end{equation}
where $\bar B_T^2 \equiv  \bar B_x^2 + \bar B_y^2$ is the ensemble average amplitude square of the transverse part of the cosmic magnetic field. In an expanding universe we can write $\bar B(\bs x, t(z))=(1+z)^2 \bar B_0(\bs x)$ where $\bar B_0(\bs x)$ is the present day amplitude of the cosmic magnetic field at the position $\bs x$. However, we do not know how the magnetic field in reality changes with the position $\bs x$ as the universe expands, so in order to bypass our ignorance about the spatial structure of the cosmic  magnetic field, is usual to assume the cosmic magnetic field to be statistically homogeneous and isotropic, see Ref. \cite{Durrer:2013pga} for details. In this case the spatial ensemble average of the magnetic field components in Fourier space\footnote{The Fourier transform of the background magnetic field is defined as $\bar{\bs B}(\bs x)=(2\pi)^{-3} \int d^3x\, \bar{\bs B}(\bar{\bs k})\, e^{i \bar{\bs k}\bs x}$ and its inverse $\bar{\bs B}(\bs k)= \int d^3\bar k\, \bar{\bs B}(\bs x)\, e^{-i \bar{\bs k}\bs x}$.} must satisfy the condition 
\begin{equation}\label{corr}
\langle \bar B_i(\bar{\bs k}) \bar B_j^*(\bar{\bs k}^\prime) \rangle = (2\pi)^3 \delta^3(\bar{\bs k} - \bar{\bs k^\prime}) \left[ (\delta_{ij} - \hat{\bar{k_i}} \hat{\bar{k_i}} ) P_{\bar{B}}(k) -i\epsilon_{ijm}\hat{\bar{k_m}} P_{\bar B}^\text{a} \right], \quad (i, j=x, y, z),
\end{equation}
where $\bar{\bs k}$ is the wave-vector associated to the background magnetic field in Fourier space, $\hat{\bar{\bs k}}= \bar{\bs k}/\bar k$, $P_{\bar B}(k)$ is the power spectrum of the cosmic magnetic field and $P_{\bar B}^\text{a}(k)$ is its anti-symmetrical part \cite{Durrer:2013pga}. One can show that the spatial  ensemble average\footnote{The maximum upper limit of integration should be up to $k\ll m_e$. However, since $m_e$ is a relatively large mass, we can formally assume the integration as extending up to infinity in the Fourier transform. } of 
\begin{equation}\label{en-dens}
\bar B_0^2 \equiv \langle \bar{\bs B_0}(\bs x) \bar{\bs B_0}(\bs x) \rangle = \frac{1}{\pi^2} \int_0^\infty dk \,k^2\, P_{\bar B}(k).
\end{equation}
By using expression \eqref{corr} and \eqref{en-dens}, one can show that $\bar B_{0, T}^2 =(2/3) \bar B_0^2 $ and consequently expression \eqref{av-velo} becomes 
\begin{equation}\label{av-velo-1}
\bar v(z)  = 1-(22/3)\rho (1+z)^4 \bar B_{0}^2.
\end{equation}
Using expression \eqref{av-velo-1} in \eqref{delta-d}, we get the following expression for $\Delta d(z)$
\begin{equation}\label{delta-d-1}
\Delta d^\text{LT}(z) = \frac{22\,\alpha \, c H_0^{-1} \bar B_0^2}{540\pi B_c^2} \int_{0}^{z}\,\frac{ (1+ z^\prime)^3 \, dz^{\prime}}{\sqrt{\Omega_\Lambda+\Omega_\text{M}(1+z^\prime)^3+\Omega_\text{R}(1+z^\prime)^4}}.
\end{equation}

Another important fact about our approximations is that we are working under the condition $\bar B\ll B_c$. This means that we need to integrate expression \eqref{delta-d-1} up to a maximum value of the redshift which is given $1+ z_\text{max} = (\kappa B_c/\bar B_0)^{1/2} $ where $0<\kappa\leq 1$. Here the factor $\kappa$ is an indicator of how much close is the strength of the cosmic magnetic field to the strength of the critical field. A value of $\kappa=1$ means that $\bar B(z)=B_c$ while $\kappa\ll 1$ means that $\bar B(z)\ll B_c$. Under such condition we are essentially requiring that  $\bar B(z)\ll B_c$ must be satisfied for all redshifts up to a maximum value $z_\text{max}$ where the approximations made so far would be valid. The biggest contribution to the integral in \eqref{delta-d-1} comes from high redshifts, so, it is quite accurate to neglect the contribution of matter and dark energy to the expression \eqref{delta-d-1} before the radiation and matter energy density equality  that happens at $1+z_\text{eq}\simeq \Omega_\text{M}/\Omega_\text{R}$, namely for $z\geq z_\text{eq}$. In this case we get from \eqref{delta-d-1}
\begin{equation}\label{delta-d-2}
\Delta d^\text{LT}(z) \simeq  \frac{11\,\alpha \, c H_0^{-1} \bar B_0^2}{540\pi \sqrt{\Omega_\text{R}}B_c^2} (z^2+2z) = 3.65\times 10^{-2} \left( \frac{\bar B_0}{\text{G}}\right)^2 (z^2+2z)  \qquad (\text{cm})  \qquad  (\text{for}\quad z_\text{eq}\lesssim z \lesssim z_\text{max})  
\end{equation}
where we used values of $cH_0^{-1}= 1.38\times 10^{28}$ cm, $\Omega_\text{R}\simeq 8.47\times 10^{-5}$ including photons and three neutrino species, and $B_c\simeq 4.41\times 10^{13}$ G. The superscript LT in the distance in \eqref{delta-d-2} indicates the light travelled distance.

Another important distance in cosmology is the comoving distance of two objects which is defined as
\begin{equation}\label{comoving-distance}
d_c^\text{CD}(z)=H_0^{-1}\int_{0}^{z}\,\frac{c\, dz^{\prime}}{\sqrt{\Omega_\Lambda+\Omega_\text{M}(1+z^\prime)^3+\Omega_\text{R}(1+z^\prime)^4}}.
\end{equation}
The comoving distance in \eqref{comoving-distance} has not a factor $1+z$ in the denominator with respect to the light travelled distance given in \eqref{light-t-distance}. In order to analytically calculate the integral appearing in \eqref{comoving-distance}, let us split the integration interval into matter and vacuum energy dominated universe and radiation energy dominated universe. For $\Omega_\Lambda+\Omega_\text{M}(1+z)^3 \gtrsim \Omega_\text{R}(1+z)^4$ and proceeding in the same way as we did above for the light travelled distance, we get the following expression for the comoving distance difference
\begin{equation}\label{delta-d-3}
\begin{gathered}
\Delta d^\text{CD}(z)  \simeq  \frac{44\,\alpha \, c H_0^{-1} \bar B_0^2}{3780\pi\,\Omega_\text{M} B_c^2} \left[\sqrt{\Omega_\text{M} + \Omega_\Lambda} \left({}_{2}{F}_1\left (1, \frac{7}{6}, \frac{5}{3}, -\frac{\Omega_\text{M}}{\Omega_\Lambda} \right ) - 1 \right) + \right.  \\ \left.  ( 1+z)^2 \sqrt{\Omega_\text{M}(1+z)^3  + \Omega_\Lambda} \left(1 -{}_{2}{F}_1\left (1, \frac{7}{6}, \frac{5}{3}, -\frac{\Omega_\text{M}(1+z)^3}{\Omega_\Lambda} \right ) \right) \right] = 6.19\times 10^{-4} \left( \frac{\bar B_0}{\text{G}}\right)^2 \left[\sqrt{0.31 + 0.68} \left({}_{2}{F}_1\left (1, \frac{7}{6}, \frac{5}{3}, -0.45\right ) -1 \right) \right.  \\ \left. + ( 1+z)^2 \sqrt{0.31(1+z)^3  + 0.68} \left(1 -{}_{2}{F}_1\left (1, \frac{7}{6}, \frac{5}{3}, -0.45(1+z)^3 \right ) \right) \right] \qquad (\text{cm}) \qquad  (\text{for}\quad 0 \lesssim z \lesssim z_\text{eq} )
\end{gathered}
\end{equation}
where ${}_{2} F_1()$ is the hypergeometric function. In the case when $\Omega_\Lambda+\Omega_\text{M}(1+z)^3 \lesssim \Omega_\text{R}(1+z)^4$, we get
\begin{equation}\label{delta-d-4}
\begin{gathered}
\Delta d^\text{CD}(z)  \simeq  \frac{22\,\alpha \, c H_0^{-1} \bar B_0^2}{1620\pi\,\sqrt{\Omega_\text{R}} B_c^2} \left[ z^3 + 3z(1+z) \right] = 2.43\times 10^{-2} \left( \frac{\bar B_0}{\text{G}}\right)^2 \left[ z^3 + 3z(1+z) \right] \qquad (\text{cm}) \qquad (\text{for}\quad z_\text{eq}\lesssim z \lesssim z_\text{max})  
\end{gathered}
\end{equation}
Clearly the difference in the comoving distance increases as we go backward in time for large values of $z$ and decreases as we approach current epoch where $\Delta d^\text{CD}\rightarrow 0$ for $z\rightarrow 0$.

Besides the comoving distance there are also two other distances that are important in cosmology which are the luminosity distance defined as $d_c^\text{L}(z) = (1+z) d_c^\text{CD}(z)$ and the angular diameter distance defined as $d_c^\text{A}(z) = d_c^\text{CD}(z)/(1+z)$. Obviously the redshift factor that multiplies the comoving distance in $d_c^\text{L}(z) $ and $d_c^\text{A}(z)$ is outside the integral appearing in the comoving distance in \eqref{comoving-distance}. This fact implies that $\Delta d^\text{L}(z)= (1+z) \Delta d^\text{CD}(z)$ and $\Delta d^\text{A}(z)= \Delta d^\text{CD}(z)/(1+z)$.  Now let us estimate the various cosmological distances discussed above with some realistic values of the parameters. Consider first the light travelled distance given in \eqref{delta-d-2} and let us take for example $z=z_\text{max}$, where $(1+z_\text{max})^2 = \kappa B_c/\bar B_0$. Then we would get from \eqref{delta-d-2} $\Delta d^\text{LT}(z_\text{max}) \simeq 1.6 \times 10^{12} \kappa (\bar B_0/\text{G})$ cm. If we take for example $\kappa=0.1$ and $\bar B_0\simeq 10^{-9}$ G, then we would get $\Delta d^\text{LT}(z_\text{max}) \simeq 160$ cm. For a magnetic field of the order $\bar B_0\simeq 10^{-7}$ G, then we would get $\Delta d^\text{LT}(z_\text{max}) \simeq 1.6\times 10^3$ cm for $\kappa =0.1$. So we see that the light traveled distance advance is in general very small even if the object that emits light is located at very high redshifts. These results apply as well to the CMB itself or any other form of electromagnetic radiation.

In the case of comoving distance difference, for example in the matter and vacuum energy dominated epoch, we would get from \eqref{delta-d-3}, $\Delta d^\text{CD}(z_\text{dec}) \simeq 2.35 \times 10^{-11}$ cm for $\bar B_0\simeq 10^{-9}$ G and $\kappa=0.1$ at the decoupling redshift $z_\text{dec}=1090$. For a magnetic field strength of three orders of magnitude larger we would get a distance advance of six orders of magnitude larger. However, the comoving distance difference is extremely small to be interesting for any source emitting at $z\lesssim z_\text{eq}$. On the other hand, for $z_\text{eq}\lesssim z \lesssim z_\text{max} $ the situation changes a lot and the comoving distance difference increases with the redshift. For example at $(1+z_\text{max})^2 = \kappa B_c/\bar B_0\gg 1$, we would get from \eqref{delta-d-4}
\begin{equation}\label{delta-d-5}
\Delta d^\text{CD}(z_\text{max})  \simeq  7.12\times 10^{18} \kappa^{3/2} \left( \frac{\bar B_0}{\text{G}}\right)^{1/2} \quad (\text{cm}).
\end{equation}
If we take for example $\kappa = 0.1$ and $\bar B_0=10^{-9}$ G, we get from expression \eqref{delta-d-5}, $\Delta d^\text{CD}(z_\text{max})  \simeq  7.12\times 10^{12}$ cm. For $\bar B_0 = 10^{-6}$ G, we get $\Delta d^\text{CD}(z_\text{max})  \simeq  2.25\times 10^{14}$ cm for $\kappa = 0.1$. So we can see that the comoving distance difference becomes larger for a source located at high redshifts.  We can use expression \eqref{delta-d-4} to calculate the luminosity distance which is given by $\Delta d^\text{L}(z)= (1+z) \Delta d^\text{CD}(z)$. 
\begin{equation}\label{delta-d-6}
\begin{gathered}
\Delta d^\text{L}(z)  \simeq 2.43\times 10^{-2} \left( \frac{\bar B_0}{\text{G}}\right)^2 \left[ (1+z)^4 - (1+z) \right] \qquad (\text{cm}) \qquad (\text{for}\quad z_\text{eq}\lesssim z \lesssim z_\text{max}).  
\end{gathered}
\end{equation}
If we take again $(1+z_\text{max})^2 = \kappa B_c/\bar B_0$, we would get from \eqref{delta-d-6}
\begin{equation}\label{delta-d-7}
\Delta d^\text{L}(z_\text{max})  \simeq 4.72 \times 10^{25}\, \kappa^2 \quad (\text{cm}).
\end{equation}
We may realize that the luminosity distance difference calculated for a source located at $z=z_\text{max}$ corresponds approximately to a distance difference $\Delta d^\text{L}(z_\text{max}) \simeq 0.15$ Mpc for $\kappa=0.1$. On the other hand the angular diameter distance difference $\Delta d^\text{A}(z)= \Delta d^\text{CD}(z)/(1+z)$ is strongly suppressed at both low and high redshifts.

It is very useful to compare the impact of non-linear QED effects to the distance measure with another phenomena that occurs in the cosmological plasma and which decreases the group velocity of electromagnetic radiation, namely the plasma oscillations\footnote{In this example, I consider the cosmological plasma to be un-magnetized and mostly composed of free electrons. The case of a cosmological magnetized plasma can be studied straightforwardly.}. Indeed, it can be easily shown that the group velocity of electromagnetic radiation due to plasma oscillations is given by
\begin{equation}
v_g^\text{pl}= c\sqrt{1 - \omega_\text{pl}^2/\omega^2}<1, \qquad \omega>\omega_\text{pl}.
\end{equation}
where $\omega_\text{pl}= \sqrt{4\pi\alpha n_e/m_e}\simeq 5.65\times 10^4 (n_e/\text{cm}^{-3})^{1/2} (\text{rad}/$s) is the plasma frequency and $n_e$ is the free electron number density. Consider for example the case of the luminosity distance difference and also consider $\omega\gg \omega_\text{pl}$. In an expanding universe one can write the free number density of electrons\cite{Ejlli:2016avx} as $n_e(z)=0.76\, n_B (1+z)^3 X_e(z)$ where $n_B\simeq 2.47 \times 10^{-7}$ cm$^{-3}$ is the present day value of baryon number density and $X_e(z)$ is the ionization fraction. In an expanding universe the relation $\omega\gg \omega_\text{pl}$ reduces to $\nu_0\gg 3.89 \sqrt{(1+z) X_e(z)}$ (Hz) where we used $\omega_0=2\pi\nu_0$. We get for the luminosity distance difference 
\begin{equation}\label{plasma-dis}
\Delta d_\text{pl}^\text{L}(z) = (1+z)\Delta d^\text{CM}(z) \simeq 7.58\, c H_0^{-1} (\text{Hz}/\nu_0)^2 (1+z) \int_{0}^{z}\,\frac{(1+z^\prime)\,X_e(z^\prime) \, dz^{\prime}}{\sqrt{\Omega_\Lambda+\Omega_\text{M}(1+z^\prime)^3+\Omega_\text{R}(1+z^\prime)^4}}.
\end{equation}
Suppose for example that we consider sources of electromagnetic radiation emitting during the vacuum energy dominated period that happens for redshifts $z< (\Omega_\Lambda/\Omega_\text{M})^{1/3}-1\simeq 0.29$. During this epoch we also have to high accuracy that $X_e(z)\simeq 1$\cite{Ejlli:2016avx}, so, we get from \eqref{plasma-dis}
\begin{equation}\label{plasma-dis-1}
\Delta d_\text{pl}^\text{L}(z) \simeq \frac{7.58\, cH_0^{-1}}{\sqrt{\Omega_\Lambda}}\left(\frac{\text{Hz}}{\nu_0}\right)^2\,(1+z) (z+z^2/2), \qquad \text{for}\qquad z<0.29\qquad \text{and} \qquad \nu_0\gg 3.89 \sqrt{1+z}\quad (\text{Hz}).
\end{equation}
If for example we consider a source emitting electromagnetic radiation at $z= 0.1$ with received frequency at $z=0$ of $\nu_0= 50$ Hz, we get from \eqref{plasma-dis-1} that the luminosity distance difference due to plasma oscillations is about $\Delta d^\text{L}(z=0.1) \simeq 1.89$ Mpc. For higher frequencies the luminosity distance difference is strongly suppressed. On the other hand for $z\gtrsim z_\text{eq}$, namely during the radiation dominated epoch we get 
\begin{equation}\label{plasma-dis-2}
\Delta d_\text{pl}^\text{L}(z) \simeq \frac{7.58\, cH_0^{-1}}{\sqrt{\Omega_\text{R}}}\left(\frac{\text{Hz}}{\nu_0}\right)^2\,(1+z) \ln(1+z), \qquad \text{for}\qquad z\gtrsim z_\text{eq}\qquad \text{and} \qquad \nu_0\gg 3.89 \sqrt{1+z}\quad (\text{Hz}).
\end{equation}
If we take for example $1+z=10^{10}$ where $X_e(z)\simeq 1$ and $\nu_0=10^7$ Hz, we would get from \eqref{plasma-dis-2}, $\Delta d^\text{L}(z=10^{10}) \simeq 8.48$ Gpc.

\section{Conclusions}
\label{sec:6}

In this work, I studied the consequences of the Euler-Heisenberg theory on the interaction of electromagnetic waves with an external magnetic field. While the consequences of this theory have been widely studied in the literature, very little attention has been paid to the appearance of the longitudinal photon state. As I have shown, in the case when the external magnetic field has a longitudinal component with respect to the direction of propagation of the electromagnetic wave, it is also generated a longitudinal electric field (or longitudinal photon). The appearance of this longitudinal state is analogous to the appearance of the longitudinal electric field in other situations in physics such as in plasma or the presence of a pseudoscalar field \cite{Ejlli:2020ifc}. 

The physical implications of the appearance of the longitudinal photon state are that in the case when the external magnetic field has a longitudinal component along the direction of propagation of the wave, it is not possible anymore to have a net separation between transverse and longitudinal electromagnetic waves. The presence of the external magnetic field and the appearance of the longitudinal electric field influence on the velocity of the two propagating modes. As I have shown in \eqref{gp-velo} the two propagating modes have different group velocities that for weak magnetic field strength compared to the critical field strength are smaller than the velocity of light in a vacuum.

Another new results that I have derived in this work are the expressions of the elements of the photon polarization tensor that are very important quantities to calculate the CMB circular polarization\cite{Ejlli:2016avx}. When the CMB propagates through cosmic magnetic fields, to study the generation of the circular polarization it is very important to have all elements of the photon polarization tensor associated with the CMB interaction with the cosmic magnetic field. The expressions of the elements of the photon polarization tensor derived in \eqref{El-Pi-1} are valid for arbitrary direction of the cosmic magnetic field with respect to the observer and have been expressed entirely in terms of the propagating wave energy. The new results found in \eqref{El-Pi-1} would allow us to study the generation of the CMB circular polarization for a completely arbitrary cosmic magnetic field direction.

The change in the group velocity of the electromagnetic radiation to that in a vacuum has consequences on distance measures in cosmology as discussed in Sec. \ref{sec:5}. As far as a cosmic magnetic field is concerned, the change in group velocity of the electromagnetic radiation in an expanding universe, would result in a slightly slower electromagnetic radiation that we receive from any source of electromagnetic radiation in our universe. Because of a slower velocity of propagation also distance measures in cosmology are slightly smaller than those obtained by using the velocity of light in a vacuum for their calculations.  As we have seen in Sec. \ref{sec:5}, bigger is the redshift of the emitting source of electromagnetic radiation, larger is the distance difference. The distance difference is very small and completely negligible for sources of low redshift such as those emitting at the post decoupling epoch. In this epoch independently of the distance measure, the distance difference is almost irrelevant for any practical purposes. It is only for those sources of electromagnetic radiation, such as primordial black holes, cosmic strings etc. located at very high redshifts that the distance difference starts becoming very important. In the case of the light traveled distance, the distance difference turns out to be very small even at redshifts when $\bar B(z)\rightarrow B_c$. This essentially implies that the difference in the age of our universe for $z\lesssim z_\text{max}$ is very small and it amounts of less than a second. However, are the comoving and luminosity distance differences that acquire large values as we approach the maximum redshift allowed by the approximations used in this work. For example the luminosity distance difference calculated at the maximum redshift $\Delta d^\text{L}(z_\text{max})$ given by \eqref{delta-d-7} is quite substantial and of the order of Mpc. I also compared the distance difference due to non-linear QED effect with the distance difference due to plasma oscillations and it turn out that the distance difference due to plasma oscillations is usually larger than that due to non-linear QED effects. In both cases, the distance differences can be quite large in terms of the absolute distance differences $\Delta d(z)$ (of the order of Mpc and Gpc) but the relative distance differences $\Delta d(z)/d_c(z)$ are very small.

The results that I have derived in this work are valid as far as the spacetime variation of the interacting electromagnetic fields are smaller than the Compton wavelength and Compton time, which essentially means that the total electromagnetic field tensor must satisfy the relation $|\partial_\mu F^{\alpha\beta}| \ll m_e |F^{\alpha\beta}|$. As we have discussed in Sec. \ref{sec:5}, these conditions translate into conditions on the energy and wavelength of the interacting electromagnetic fields where we must have $\omega\ll m_e$ for the propagating electromagnetic field. Another approximation used in this work has been that we worked in the regime when $\bar B(z)\ll B_c$. One might ask what happens to our results about changes in the group velocities and distance differences discussed in Sec. \ref{sec:5} in the case we are in the regime when $\bar B(z)\gg B_c$ and the Euler-Heisenberg theory is studied at the temperature of $T\neq 0$ such is the case of the early universe? The answers to these questions are not easy since one needs to start with the general expression given in \eqref{EH-lan} and study the consequences of the theory at the temperature of $T\neq 0$. These studies are beyond the scope of this work and will be addressed in the future elsewhere.

\end{document}